\documentclass[twocolumn, amsmath,amssymb, aps,10pt]{revtex4}

\usepackage{graphicx,bm,color}

\usepackage{braket,hyperref}

\begin{document}

\preprint{APS/123-QED}

\title{Quantum Interference in Time-Delayed Nonsequential Double Ionization}%

\author{A. S. Maxwell}
	\email{andrew.maxwell.14@ucl.ac.uk}

\author{C. Figueira de Morisson Faria}%
 \email{c.faria@ucl.ac.uk}
\affiliation{%
Department of Physics \& Astronomy, University College London \\Gower Street  London  WC1E 6BT, United Kingdom
}%

\date{\today}
            
\begin{abstract}
We perform a systematic analysis of quantum interference in nonsequential double ionization focusing on the recollision-excitation with subsequent ionization (RESI) mechanism, employing the strong-field approximation (SFA). We find that interference has a major influence on the shape, localization and symmetry of the correlated electron momentum distributions. In particular, the fourfold symmetry with regard to the parallel momentum components observed in previous SFA studies is broken. Two types of interference are observed and thoroughly analyzed, namely that caused by electron indistinguishability and intra-cycle events, and that stemming from different excitation channels. We find that interference is most prominent around the diagonal and anti-diagonal in the parallel-momentum plane and provide fully analytical expressions for most interference patterns encountered. We also show that this interference can be controlled by an appropriate choice of phase and excited-state geometry. This leads a to myriad of shapes for the RESI distributions including correlated, anti-correlated and ring-shaped.

\end{abstract}

\maketitle

\section{\label{sec1:level1}Introduction}
Quantum interference in strong-field phenomena has established itself as a very powerful attosecond-imaging tool.  Well-known examples of interference are dynamic and structural patterns in molecular high-order harmonic generation (HHG) (for reviews see, e.g., \cite{Lein2007} and our recent publication \cite{Augstein2012}), and the fan-shaped structure in above-threshold ionization (ATI) of rare-gas atoms \cite{Rudenko2004a,Chen2006,Arbo2006a,Arbo2008,Yan2012}. The connection with imaging stems from the physical mechanism behind these phenomena, namely the laser-induced recollision or recombination of an electron with its parent ion \cite{Corkum1993}. Recombination and elastic recollision lead to HHG and high-order ATI, respectively. If the electron is released in the continuum and reaches the detector without further interaction, this will lead to direct ATI. Thus, HHG and ATI transition amplitudes may be associated with electron orbits and there will be many possible routes for the active electron. Hence, quantum mechanically, they will interfere. 

Strong-field nonsequential double ionization (NSDI) may also be described as the recollision of an electron with its parent ion. However, upon return the electron gives part of its kinetic energy to the core. This leads to the release of a second electron. Hence, a legitimate question is whether quantum interference also influences NSDI. There are, however remarkably fewer studies of interference effects in this context, even though NSDI has been investigated for over two decades.

This apparent lack of interest may be related to the widespread belief that interference does not play a significant role in NSDI. This may be attributed to the success of classical NSDI models (for reviews see \cite{Faria2011,Becker2012}), which have reproduced key features in NDSI electron-momentum distributions and have shown a very good agreement with experimental findings. These include, for instance, the shapes and maxima of the electron-momentum distributions, and even finer details such as the v-shaped structure that is a fingerprint of the long-range electron-electron interaction \cite{Ye2008,Emmanouilidou2008}. Furthermore, several studies of quantum-classical correspondence in NSDI have suggested that interference  will not survive integration over momentum components perpendicular to the laser-field polarization. This is the typical scenario in NSDI experiments. Such conclusions have been inferred from the excellent agreement between classical models and the full solution of the time-dependent Schr\"odinger equation \cite{Panfili2001}, or the strong-field approximation (SFA) \cite{FigueiradeMorissonFaria2004a,Faria2004,Jia2013}.

The above-stated studies, however, have focused on the electron impact ionization mechanism, in which the first electron, by re-colliding with the core, immediately releases a second electron. Another physical mechanism, which is less studied, is re-collision-excitation with subsequent ionization (RESI). In RESI, the first electron does not provide the second electron with enough energy to be released in the continuum. Hence, it is excited to another bound state, from which it subsequently leaves via tunneling ionization. 

The prevalent view is that the above-mentioned time delay leads to anti-correlated electron momentum distributions. They would populate the second and fourth quadrant of the plane spanned by the electron momentum components $p_{n\parallel}$, $(n=1,2)$ parallel to the driving-field polarization. These features have been identified in experiments performed in the below-threshold intensity regime \cite{Weber2000,Eremina2003,Zeidler2005,Liu2008,Liu2010}. However, more recent results strongly suggest that this interpretation is over-simplified. 

For instance, recent experiments of RESI of Ar with few-cycle pulses have revealed cross-shaped distributions strongly localized along the axes $p_{n\parallel}=0$ \cite{Bergues2012}. Subsequent studies by the same group have shown that if the pulse length is increased, the distributions spread across the four quadrants of the parallel momentum plane with a slight preference for back-to-back emission \cite{Kubel2014}. This agrees with the findings in \cite{Sun2014} for NSDI in Xe, namely distributions equally occupying all momentum quadrants that exhibited RESI characteristics. Hence, it is plausible that in \cite{Eremina2003,Zeidler2005,Liu2008,Liu2010}, the signal in the first and third quadrant of the parallel momentum plane, which was dismissed as electron-impact ionization, could in fact be RESI.  

This affirmative is backed by theoretical findings using methods as diverse as the SFA and related methods \cite{Shaaran2010,Shaaran2010a,Chen2010} and classical-trajectory \cite{Emmanouilidou2009,Ye2010,Sun2014,Zhang2014} computations. Although back-to-back emission was highlighted, in many classical-trajectory studies cross- or ring-shaped distributions spreading across all quadrants have been identified \cite{Emmanouilidou2009,Ye2010,Sun2014,Zhang2014}. This behavior has even been found for intensities far below the threshold, for which electron-impact ionization can definitely be ruled out \cite{Ye2010}.  In particular for semi-analytical methods such as the SFA \cite{Shaaran2010,Shaaran2010a,Shaaran2011} or the quantitative rescattering theory \cite{Chen2010}, exclusively fourfold symmetry distributions were identified for RESI. One should note, however, that in none of the SFA computations has interference between different events or excitation channels been incorporated. Recently, however, SFA computations using inter-channel interference have shown that the fourfold symmetry can be broken \cite{Hao2014}. Indeed, it has been argued that quantum interference is paramount, and that it may lead to anti-correlated distributions. 

Nonetheless, more systematic studies of quantum-interference effects in RESI are missing. This interference may  occur (a) between events which are displaced by half a cycle and those present due to the symmetry of indistinguishable electrons, (b) between different channels of excitation for the second electron and (c) between different orbits along which the first electron may return.

In this article, we present a systematic analysis of the two first types of interference, employing the SFA.  The SFA, if used in conjunction with the steepest descent method, provides a very intuitive interpretation in terms of electron orbits, and retains quantum interference and tunneling. This makes it an ideal tool for analyzing different types of interference. The last type of interference is incorporated in the model, but is washed out when the perpendicular momentum components are integrated over. This has been studied in previous publications \cite{Shaaran2010,Shaaran2012}. 

This work is organized as follows. In Sec.~\ref{sec2:level1} we briefly review the necessary background for understanding the subsequent results. These are provided in Secs. \ref{sec3:level1}, \ref{sec4:level1} and \ref{sec5:level1}. In Sec.~\ref{sec3:level1}, we investigate symmetry-related interference and provide expressions for the features encountered. Subsequently, in Sec.~\ref{sec4:level1}, we analyze how the geometry of the bound states involved modifies this interference, and  in Sec.~\ref{sec5:level1}, we study how different excitation channels interfere and how this affects the electron-momentum distributions.  Finally, in Sec.~\ref{sec6:level1}, we provide an overall discussion and state our main conclusions.
\section{\label{sec2:level1}Background}
\subsection{Transition amplitude and saddle-point equations}

Within the strong-field approximation and in atomic units, the RESI transition amplitude reads
\begin{eqnarray}
&&M(\mathbf{p}_{1},\mathbf{p}_{2})=\hspace*{-0.2cm}\int_{-\infty }^{\infty
}dt\int_{-\infty }^{t}dt^{^{\prime }}\int_{-\infty }^{t^{\prime
}}dt^{^{\prime \prime }}\int d^{3}k  \notag \\
&&\times V_{\mathbf{p}_{2}e}V_{\mathbf{p}_{1}e,\mathbf{k}g}V_{\mathbf{k}%
g}\exp [iS(\mathbf{p}_{1},\mathbf{p}_{2},\mathbf{k},t,t^{\prime },t^{\prime
\prime })],  \label{eq:Mp}
\end{eqnarray}%
where
\begin{eqnarray}
&&S(\mathbf{p}_{1},\mathbf{p}_{2},\mathbf{k},t,t^{\prime },t^{\prime \prime
})=  \notag \\
&&\quad E_{\mathrm{1g}}t^{\prime \prime }+E_{\mathrm{2g}}t^{\prime
}+E_{\mathrm{2e}}(t-t^{\prime })-\int_{t^{\prime \prime }}^{t^{\prime }}%
\hspace{-0.1cm}\frac{[\mathbf{k}+\mathbf{A}(\tau )]^{2}}{2}d\tau  \notag \\
&&\quad -\int_{t^{\prime }}^{\infty }\hspace{-0.1cm}\frac{[\mathbf{p}_{1}+%
\mathbf{A}(\tau )]^{2}}{2}d\tau -\int_{t}^{\infty }\hspace{-0.1cm}\frac{[%
\mathbf{p}_{2}+\mathbf{A}(\tau )]^{2}}{2}d\tau  \label{eq:singlecS}
\end{eqnarray} gives the semiclassical action, and the prefactors
\begin{eqnarray}
V_{\mathbf{k}g} &=&\left\langle \tilde{\mathbf{k}}(t^{\prime \prime
})\right\vert V\left\vert \psi _{1}^{(g)}\right\rangle  \notag \\
&=&\frac{1}{(2\pi )^{3/2}}\int d^{3}r_{1}V(\mathbf{r}_{1})e^{-i\tilde{\mathbf{%
k}}(t^{\prime \prime })\cdot \mathbf{r}_{1}}\psi _{1}^{(g)}(\mathbf{r}%
_{1}),  \label{eq:Vkg}
\end{eqnarray}%
\begin{eqnarray}
V_{\mathbf{p}_{1}e,\mathbf{k}g} &=&\left\langle \tilde{\mathbf{p}}_{1}\left(
t^{\prime }\right) ,\psi _{2}^{(e)}\right\vert V_{12}\left\vert \tilde{\mathbf{\
k}}(t^{\prime }),\psi _{2}^{(g)}\right\rangle  \notag \\
&=&\frac{1}{(2\pi )^{3}}\int \int d^{3}r_{2}d^{3}r_{1}\exp [-i(\mathbf{p}%
_{1}-\mathbf{k})\cdot \mathbf{r}_{1}]  \notag \\
&&{}\times V_{12}(\mathbf{r}_{1,}\mathbf{r}_{2})[\psi _{2}^{(e)}(\mathbf{r}%
_{2})]^{\ast }\psi _{2}^{(g)}(\mathbf{r}_{2})\quad  \label{eq:Vp1e,kg}
\end{eqnarray}
and
\begin{eqnarray}
V_{\mathbf{p}_{2}e} &=&\left\langle \tilde{\mathbf{p}}_{2}\left( t\right)
\left\vert V_{\mathrm{ion}}\right\vert \psi _{2}^{(e)}\right\rangle  \notag
\\
&=&\frac{1}{(2\pi )^{3/2}}\int d^{3}r_{2}V_{\mathrm{ion}}(\mathbf{r}%
_{2})e^{-i\tilde{\mathbf{p}}_{2}(t)\cdot \mathbf{r}_{2}}\psi _{2}^{(e)}(%
\mathbf{r}_{2})  \label{eq:Vp2e}
\end{eqnarray}%
incorporate all information about the interactions and electronic bound states. Specifically, Eqs.~(\ref{eq:Vkg}), (\ref{eq:Vp1e,kg}) and (\ref{eq:Vp2e}) are related to, the ionization of the first electron, the recollision of the first electron with excitation of the second electron, and the tunnel ionization of the second electron, respectively. Therein,  $V(\mathbf{r}_{1})$ and $V_{\mathrm{ion}}(\mathbf{r}_{2})$ denotes the binding potential ``seen" by
the first and the second electron, respectively, and  $V_{12}(\mathbf{r}_{1},\mathbf{r}_{2})$ gives the electron-electron interaction.  Furthermore, $\tilde{\mathbf{k}}(\tau )=\mathbf{k}+\mathbf{A}(\tau )$
and $\tilde{\mathbf{p}}_{n}(\tau )=\mathbf{p}_{n}+\mathbf{A}(\tau )$
$(\tau =t,t^{\prime},t^{\prime \prime }$) in the length gauge, and
$\tilde{\mathbf{k}}(\tau )=\mathbf{k}$ and
$\tilde{\mathbf{p}}_{n}(\tau )=\mathbf{p}_{n}$ in the velocity
gauge, with $n=1,2$. In our previous publication \cite{Shaaran2010}, we have verified that, in practice, the
results obtained in both gauges lead to qualitatively similar results. Here we employ the latter gauge in order to avoid bound-state singularities. 

The transition amplitude (\ref{eq:Mp}) describes a process in which the first electron,
initially bound in the ground state $|\psi _{1}^{(g)}\rangle $, is released at a time
$t^{\prime \prime }$ into a continuum state, which is approximated by the Volkov state $|\tilde{\mathbf{k}}
(t^{\prime \prime })\rangle $. Subsequently, it remains in the continuum from the time $t^{\prime \prime }$ to the time $t^{\prime
}$ with intermediate momentum $\mathbf{k}$. At $t^{\prime }$,
it returns to its parent ion and interacts with a core electron via $V_{12}$. This interaction excites the second electron
from the ground state $|\psi _{2}^{(g)}\rangle $ of the singly ionized target to the
state $|\psi _{2}^{(e)}\rangle$.  The first electron reaches the detector
with final momentum $\mathbf{p}_{1}$ immediately after rescattering.
The second electron remains bound until a later time $t$, when it is
released by tunnel ionization into a Volkov state
$|\tilde{\mathbf{p}}_{2}\left( t\right) \rangle $. It reaches the
detector with final momentum $\mathbf{p}_{2}$.  The ground-state energy of the neutral system is given by  $E_{\mathrm{1g}}$, and the energies of the ground and excited states of the singly ionized target are $E_{\mathrm{2g}}$ and $E_{\mathrm{2e}}$, respectively. For details on how this transition amplitude is derived we refer to our previous publication \cite{Shaaran2010a}.

Throughout, we employ the steepest descent method in order to compute the transition amplitude (\ref{eq:Mp}). In this method, we seek  variables $t$, $t^{\prime}$ $t^{\prime\prime}$ and $\mathbf{k}$ so that the action is stationary. This leads to the saddle-point equations
\begin{equation}
\left[ \mathbf{k}+\mathbf{A}(t^{\prime \prime })\right] ^{2}=-2E_{\mathrm{1g}},
\label{eq:saddle1}
\end{equation}%
\begin{equation}
\mathbf{k=}-\frac{1}{t^{\prime }-t^{\prime \prime }}\int_{t^{\prime \prime
}}^{t^{\prime }}d\tau \mathbf{A}(\tau ),  \label{eq:saddle2}
\end{equation}%
\begin{equation}
\left[ \mathbf{p}_{1}+\mathbf{A}(t^{\prime })^2 \right]=\left[ \mathbf{k}+\mathbf{A}%
(t^{\prime })\right] ^{2}-2(E_{\mathrm{2g}}-E_{\mathrm{2e}}), \label{eq:saddle3}
\end{equation}
and
\begin{equation}
\lbrack \mathbf{p}_{2}+\mathbf{A}(t)]^{2}=\mathbf{-}2E_{\mathrm{2e}}.
\label{eq:saddle4}
\end{equation}
 Eqs.~(\ref{eq:saddle1}) and (\ref{eq:saddle4}) give the energy conservation of the first and second electron at the instant of tunnel ionization. For the former, this occurs from the ground state at the time $t^{\prime\prime}$, while the latter tunnels from an excited state at a later time $t$. Neither has a real solution, which reflects the fact that tunneling has no classical counterpart. Eq.~(\ref{eq:saddle2}) restricts the intermediate momentum of the first electron so that it returns to the core, which is assumed to be located at the origin. Finally, Eq.~(\ref{eq:saddle3}) states that, upon return, the first electron gives part of its kinetic energy upon return to ``bridge" the gap $E_{\mathrm{2g}}-E_{\mathrm{2e}}$ and promotes the second electron to an excited state. We use both the standard saddle point approximation and a uniform asymptotic expansion whose only applicability requirement is that the orbits occur in pairs. For details on these methods see our previous work \cite{Faria2002}.

\subsection{Momentum constraints}
 For simplicity, we will consider a monochromatic field throughout. Hence, the vector potential reads
 \begin{equation}
 \mathbf{A}(\tau)=2\sqrt{U_p}\cos(\omega \tau)\hat{e}_{\parallel},
\end{equation}
where $\tau$ is a generic time that may be $t,t^{\prime}$ or $t^{\prime\prime}$, $\omega$ is the driving-field frequency and $U_p=I/(4\omega^2)$ is the ponderomotive energy, which is proportional to the intensity $I$ of the driving field. This choice of field means that $A(\tau \pm T/2)=- A(\tau)$, where $T=2\pi/\omega$ denotes the field cycle. Thus, events whose times are displaced by half a cycle are related by momentum inversion.
Furthermore, since both electrons are indistinguishable, one must also exchange $\mathbf{p}_1$ and $\mathbf{p}_2$ in Eq.~(\ref{eq:Mp}) and add the corresponding amplitudes.

The distributions will be located around $(\mathbf{p}_1,\mathbf{p}_2)=(\pm 2\sqrt{U_p},0)$. This comes from the fact that the rescattering of the first electron and ionization of the second electron occur most probably near field crossings and crests, respectively. 

Estimates for the regions in the parallel momentum plane to be populated follow from the saddle-point equations. For the first electron, Eq.~(\ref{eq:saddle3}) gives
\begin{equation}
\pm 2\sqrt{U_p}-\sqrt{2\triangle E}\leq p_{1\parallel }\leq \pm 2\sqrt{U_p}+\sqrt{2\triangle E} ,\label{regionp1}
\end{equation}
 where $\triangle E$ = $E_{\mathrm{kin}}(t^{\prime },t^{\prime \prime })-%
\tilde{E}_{\mathrm{exc}}$ yields the energy difference between the
kinetic energy $\ E_{\mathrm{kin}}(t^{\prime
},t^{\prime\prime })$ of the first electron upon return and the
energy $\tilde{E}_{\mathrm{exc}}=E_{\mathrm{2g}}-E_{\mathrm{2e}}+p_{1\perp}^2/2$.  The above-stated inequality indicates that the region where rescattering has a classical counterpart, which is largest if $\mathbf{p}_{1\perp}=0$.
For the second electron, one must bear in mind that Eq.~(\ref{eq:saddle4}) is formally identical to that describing tunnel ionization in direct ATI. The direct ATI cutoff energy is $2U_p$, so that 
\begin{equation}
-2\sqrt{U_p}\leq p_{2\parallel}\leq 2\sqrt{U_p}\text{.}
\label{eq:constraint2nd}
 \end{equation}
 In this latter estimate, we have considered $\mathbf{p}_{2\perp}=0$. Eqs.~(\ref{regionp1}) and (\ref{eq:constraint2nd}) give cross-shaped electron momentum distributions strongly located around the axes of the $p_{1\parallel}p_{2\parallel}$ plane. Detailed explanations of these constraints have been provided elsewhere \cite{Shaaran2010,Shaaran2010a,Faria2012}.

\section{\label{sec3:level1}Interference of Events}
\begin{figure}
	\includegraphics[width=8cm]{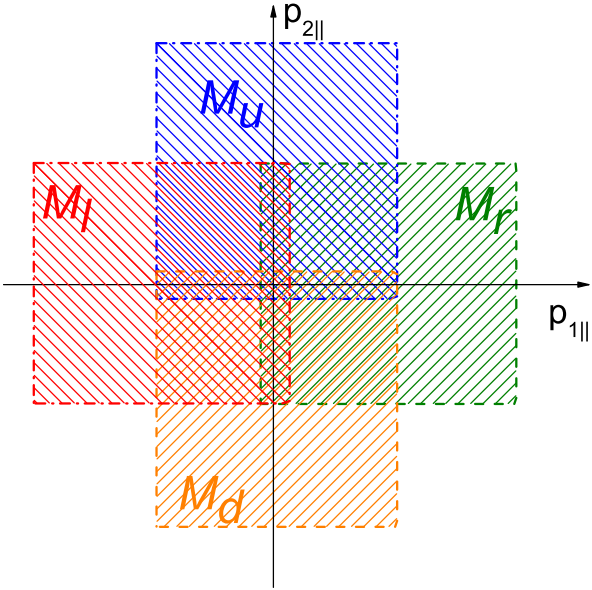}
	\caption{Schematic representation of the momentum regions occupied by the transition amplitudes $M_l$, $M_u$, $M_d$ and $M_r$, which are displayed as the rectangular patterned regions. The overlap regions indicate areas within these constraints for which quantum interference may occur. The intensity represented in the figure is high enough to allow some interference at the origin $(p_{1\parallel},p_{2\parallel})=(0,0)$.}
	\label{fig:schematicinterf}
\end{figure}

Here we will analyze interference between events displaced by half a cycle and those present due to the particle exchange symmetry of the system. This leads to four transition amplitudes, $M(\mathbf{p}_1,\mathbf{p}_2)$, $M(\mathbf{p}_2,\mathbf{p}_1)$, $M(-\mathbf{p}_1,-\mathbf{p}_2)$ and $M(-\mathbf{p}_2,-\mathbf{p}_1)$, which must be combined. Due to the localization of these transition amplitudes near the negative $p_{1\parallel}$ half axis, positive $p_{2\parallel}$ half  axis, positive $p_{1\parallel}$ and negative $p_{2\parallel}$ half axis, i.e., occupying the left, upper, right and lower regions in the parallel momentum plane, we relabel them $M_l$, $M_u$, $M_r$ and $M_d$, respectively. A schematic representation of the momentum regions occupied by the different transition amplitudes is provided in Fig.~\ref{fig:schematicinterf}.

Throughout this analysis we will compare coherent and incoherent sums of these amplitudes integrated over momentum components perpendicular to the laser field, which are given by 
\begin{equation}
\hspace*{-0.2cm}W(p_{1 \parallel},p_{2 \parallel})= \hspace*{-0.2cm}\int d^2\bm{p}_{1 \perp} d^2\bm{p}_{2 \perp} |M_l+M_u+M_r+M_d|^2
\label{coherent}
\end{equation}
and
\begin{eqnarray}
\hspace*{-0.52cm}W(p_{1 \parallel},p_{2 \parallel})&=&\hspace*{-0.2cm}\int d^2\bm{p}_{1 \perp} d^2\bm{p}_{2 \perp} \nonumber \\ &\times&\left(|M_l|^2+|M_u|^2+|M_r|^2+|M_d|^2 \right),
\label{incoherent}
\end{eqnarray}respectively.

Quantum interference occurs predominantly in the overlap regions in Fig.~\ref{fig:schematicinterf}. Apart from the region around $(p_{1\parallel},p_{2\parallel})=(0,0)$, in which, potentially, all amplitudes may interfere, due to the constraints discussed in the previous section, we expect that $M_l$ and $M_u$ will interfere predominantly in the second quadrant and that $M_l$ and $M_d$ will overlap in the third quadrant of the $p_{1\parallel}p_{2\parallel}$ plane. Similarly, interference between $M_r$ and $M_u$ is expected to occur in the first quadrant, and between $M_r$ and $M_l$ will take place mostly in the fourth quadrant. For simplicity, throughout this section we will neglect the prefactors in Eqs.~(\ref{coherent}) and (\ref{incoherent}). This will help us identify how the phases determined by the corresponding actions interact without further momentum bias.  The prefactors will be reintroduced in Sec.~\ref{sec5:level1}.

\begin{figure}

\centering
\includegraphics[width=\linewidth]{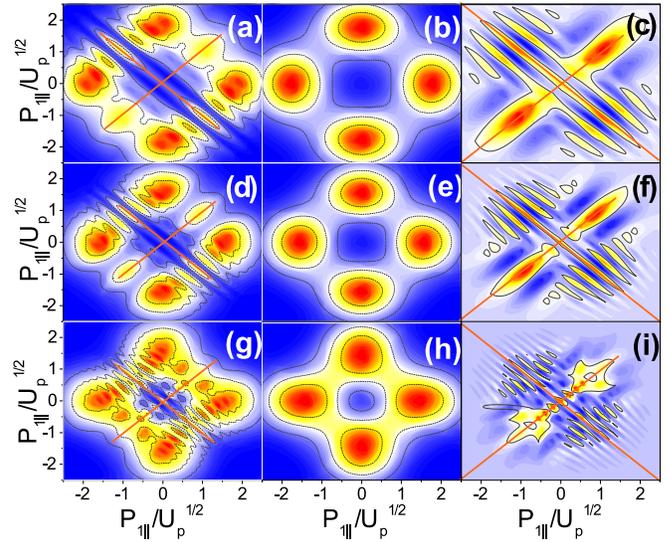}
\caption{Coherent and incoherent sums of all amplitudes integrated over the perpendicular momenta as given by Eq.~(\ref{coherent}) and (\ref{incoherent}). The columns from left to right show a coherent sum of amplitudes [panels (a), (d) and (g)], an incoherent sum of amplitudes [panels (b), (e) and (h)], and the difference between the two [panels (c), (f) and (i)]. The rows show different laser intensities  of $I=$ 2.28, 4.56 and 6.84 ($\times 10^{13}$ $\mathrm{W/cm}^2$) from top to bottom with values for $U_p$ of 0.05, 0.1 and 0.15 a.u. corresponding to an angular frequency $\omega=0.057$ a.u. The RESI channel corresponds to a ground state valence orbital $3s$ and an excited orbital $3p$ for the second electrons. The ionization potentials of the ground state of the first and second electron and the excited state of the second electron are $I_{1 g}= 0.58$, $I_{2 g}=1.02$ and $I_{2 e}=0.52$ a.u. respectively. The diagonal and anti-diagonals $p_{1\parallel}=\pm p_{2\parallel}$ are indicated with the orange lines in the figure. The signal in each panel has been normalized with regard to its maximum.}
\label{fig:FullMaps}
\end{figure}

In Fig.~\ref{fig:FullMaps}, we display the coherent and incoherent sum for three driving-field intensities. The figure shows that the interference between different events survives the integration over the transverse momentum coordinates, as there are obvious differences between coherent and incoherent sums of events.  Clearer features can be outlined from the difference of the two probability maps. There are maxima along the diagonal and anti-diagonal at all intensities and hyperbolic fringes whose presence becomes more obvious as the intensity increases. For higher intensities the patterns become more complicated.

These features can be explained by looking at the integrand of the coherent sum, which can be rewritten in terms of the actions $S_l$, $S_r$, $S_u$ and $S_d$ associated with the above-stated amplitudes. A common factor can be taken out, leaving terms that will contribute to the interference.
 Explicitly,
\begin{eqnarray}
\Omega(p_{1\parallel},p_{2\parallel})\hspace*{-0.2cm}&=&\hspace*{-0.2cm}\int\hspace*{-0.1cm} d^4p_{\perp}\left| \int\hspace*{-0.1cm} d^3 t (e^{i S_l} + e^{i S_r}+ e^{i S_u}+e^{i S_d}) \right|^2 \\
\hspace*{-0.2cm}&=&\hspace*{-0.2cm}\int\hspace*{-0.1cm}d^4 p_{\perp}\left| \int\hspace*{-0.1cm} d^3 t e^{i S_{l}}(1 + e^{i \alpha_{lr}}+ e^{i \alpha_{lu}}+e^{i \alpha_{ld}}) \right|^2,\nonumber
\label{interftotal}
\end{eqnarray}
where the action $S_l=S(\mathbf{p}_1,\mathbf{p}_2,\mathbf{k},t,t^{\prime},t^{\prime\prime})$ is associated with the matrix element $M_l=M(\mathbf{p}_1,\mathbf{p}_2)$ giving the left peak. The integrals over time and momenta have been abbreviated as
\begin{equation}
 \int d^3t=\int_{-\infty }^{\infty
}dt\int_{-\infty }^{t}dt^{^{\prime }}\int_{-\infty }^{t^{\prime
}}dt^{^{\prime \prime }},
\end{equation}
and
\begin{equation}
\int d^4p_{\perp}=\int\int d^2 p_{1\perp}d^2 p_{2\perp},
\end{equation}
and the phase differences between the actions read
\begin{align}
\alpha_{ld}&=\frac{1}{2}  \left(p_{1 }^2 -p_{2}^2\right)(t-t') \nonumber \\
&+\frac{2\sqrt{U_p}}{\omega} (p_{1 \parallel}-p_{2 \parallel}) (\sin ( \omega t)-  \sin (\omega t')),
\label{phaseleftdown}
\end{align}
\begin{align}
\alpha_{lu}&=\frac{\pi}{2 \omega }  \left(4U_p+2 \text{E}_{2 e}+2 \text{E}_{1 g}+p_{1 }^2+p_{2 }^2\right) \nonumber\\
&+\frac{1}{2}  \left(p_{1 }^2 -p_{2}^2\right)(t-t') \nonumber \\
&-\frac{2\sqrt{U_p}}{\omega} (p_{1 \parallel}+p_{2 \parallel}) (\sin ( \omega t)-  \sin (\omega t'))
\label{phaseleftup}
\end{align}
and
\begin{align}
\alpha_{lr}&=\frac{\pi}{2 \omega }  \left(4U_p+2 \text{E}_{2 e}+2 \text{E}_{1 g}+p_{1 }^2+p_{2 }^2\right) \nonumber\\
&-\frac{4\sqrt{U_p}}{\omega} \left(p_{1 \parallel} \sin (\omega t'  )+ p_{2 \parallel} \sin (\omega t  )\right).
\label{phaseleftright}
\end{align}

In Eq.~(\ref{phaseleftdown})-(\ref{phaseleftright}), the general form $\alpha_{ij}$ has been adopted, where the indices $i,j$ refer to the interfering amplitudes. Constructive interference requires that the integrand in Eq.~(\ref{interftotal}) is maximized, which will occur when $\alpha_{ij}=0$, or as small as possible. We will start by investigating the ``left-down" phase difference (\ref{phaseleftdown}), between the actions associated with the left and lower peak. This phase vanishes for arbitrary times $t,t^{\prime}$ if conditions
\begin{minipage}{4.2cm}
\begin{equation}
 p^2_1-p^2_2=0
 \label{interfalphald1}
\end{equation}
\vspace*{0.05cm}
\end{minipage}
\begin{minipage}{4.2cm}
\begin{equation}
\textrm{and}\hspace*{0.5cm} p_{1\parallel}=p_{2\parallel}
 \label{interfalphald2}
\end{equation}
\vspace*{0.05cm}
\end{minipage}
are satisfied.

Condition (\ref{interfalphald1}), if written as a function of the parallel and perpendicular momentum components for $p^2_{2\perp}-p_{1\perp}\neq 0$, give the hyperbolae
\begin{equation}
\frac{p^2_{1\parallel}}{p^2_{2\perp}-p_{1\perp}}-\frac{p^2_{2\parallel}}{p^2_{2\perp}-p_{1\perp}}=1
\label{hyperbolae}
\end{equation}
whose asymptotes lie at the diagonal and anti-diagonal $p_{1\parallel}=\pm p_{2\parallel}$ and whose vertices and transverse axis will depend on whether $p^2_{2\perp}-p^2_{1\perp}$ are positive or negative. The former and latter case will lead to hyperbolae with transverse axes along $p_{1\parallel}$ and $p_{2\parallel}$, respectively.  For equal transverse momenta, instead, condition (\ref{interfalphald1}) will give $p_{1\parallel}=\pm p_{2\parallel}$, i.e., the diagonal and the anti-diagonal. In this case, the interference condition is independent of the transverse momenta, so that they are expected to survive when the integration over these variables is performed. The hyperbolae, on the other hand, depend on the transverse momentum coordinates, but may survive integration. If this happens, however, integration may influence their transverse axes, vertices and foci.
\begin{figure}

\centering
\includegraphics[width=\linewidth]{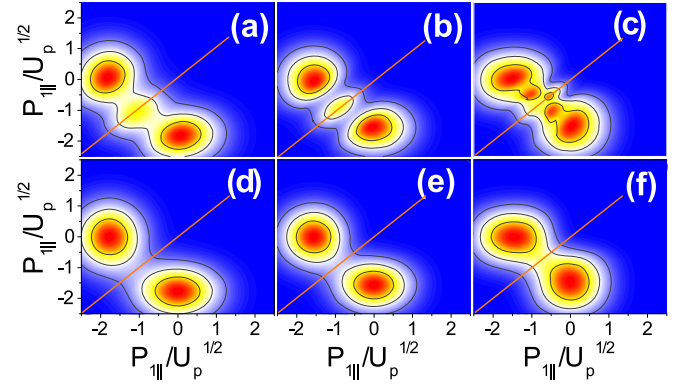}

\caption{Correlated electron-momentum distributions obtained by combining the transition amplitudes $M_l$ and $M_d$, isolating the effect of $\alpha_{l d}$, integrated over the transverse-momentum components. The left, middle and right columns have been computed for laser intensities  of $I=$ 2.28, 4.56 and 6.84 ($\times 10^{13}$ $\mathrm{W/cm}^2$), with values for $U_p$ of 0.05, 0.1 and 0.15 a.u., respectively.
Panels (a), (b) and (c) show the coherent sum $\Omega^{ud}_{coh}=|M_l+M_d|^2$, while in panels (d), (e) and (f) the incoherent sum $\Omega^{ud}_{in}=|M_l|^2+|M_d|^2$ is displayed.  The intensities and ionization potentials are the same as in Fig. \ref{fig:FullMaps}. The diagonals $p_{1\parallel}= p_{2\parallel}$ are indicated with the orange lines in the figure. The signal in each panel has been normalized with regard to its maximum. }
\label{fig:LeftandDownMaps}
\end{figure}

The analysis performed above suggests that there will be maxima along the diagonal and the anti-diagonal, and that there could be hyperbolic fringes in the coherent sum of the two-electron transition amplitudes, in agreement with Fig.~{\ref{fig:FullMaps}}. In Fig.~\ref{fig:LeftandDownMaps}, we have a closer look at this interference, and plot a partial distribution in which only $M_l$ and $M_d$ summed, coherently and incoherently (upper and lower panels, respectively). The strongest feature in the figure is the maximum along the diagonal, which comes from condition (\ref{interfalphald2}) and also from the case $p_{1\perp}=p_{2\perp}$ related to the hyperbolic condition (\ref{interfalphald1}). Parallel to the diagonal, there are also interference fringes, whose number increases with the driving-field intensity. The interference maxima along the anti-diagonal cannot be seen as the partial sum employed in the figure is vanishingly small in the second quadrant of the parallel momentum plane.

An estimate for the position of the fringes can be obtained by considering the coherent superposition of $M_l$ and $M_d$, and expanding the momenta in the vicinity of the diagonal, i.e., $p_{1\parallel}=-p_{2\parallel}+\delta$. Fringes will occur for $\exp[i\alpha_{ld}]=\pm 1$, i.e., for $\alpha_{ld}=n\pi$, where even and odd $n$ give maxima and minima, respectively. Assuming small momenta $\mathbf{p}_{n} (n=1,2)$, rescattering times at field crossings and ionization times at the subsequent field crest [$t'=n\pi/\omega$ and $t=(2n+1)\pi/(2\omega)$], the fringe spacing can be approximated as
\begin{equation}
|\delta|\simeq \frac{\omega n \pi}{2\sqrt{U_p}}.\label{Eq:fringes}
\end{equation}
The above-stated equation shows that the spacing between the fringes is inversely proportional to the driving-field strength, in agreement with what has been observed in the previous figures.
\begin{figure}
\centering
\includegraphics[width=\linewidth]{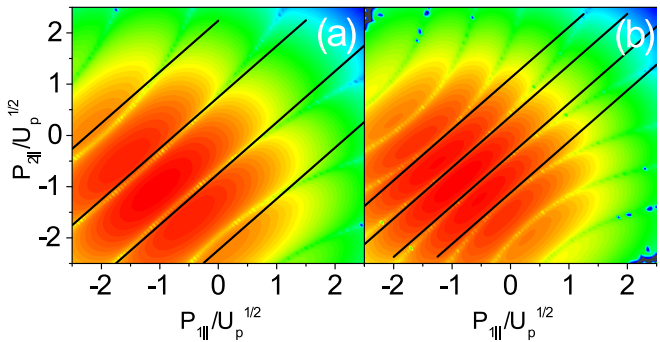}
\caption{Absolute value of the difference between the sums from the upper and lower panels in Fig. \ref{fig:LeftandDownMaps}. Panels (a) and (b) correspond to $U_p=0.05$ and $U_p=0.1$ a.u respectively. The approximate fringes given by Eq.~(\ref{Eq:fringes}) are marked by black lines.  The panels have been plotted in a logarithmic scale ($\log(|\Omega^{ud}_{coh}-\Omega^{ud}_{in}|)$). }
\label{fig:diffleftdown}
\end{figure}

The interference patterns are highlighted in Fig.~\ref{fig:diffleftdown}, where we display the difference between the coherent and the incoherent sum, for the two lower driving-field intensities in the previous figure. Overall, for small momenta the fringe spacing exhibits a very good agreement with Eq.~(\ref{Eq:fringes}). Furthermore, all panels in the figure exhibit clear hyperbolic structures, whose number increases with the driving-field intensity. One should note that their transverse axis is not located along $p_{n\parallel}=0$ $(n=1,2)$. This displacement is probably related to the integration over the transverse momenta, which influence the direction of the hyperbolae. Furthermore, the last diagonal term will act to shift the center of the hyperbola along the diagonal, which can be observed by the fact the hyperbolae are opening, instead of exhibiting  asymptotic behavior towards the diagonals. As the laser intensity increases the hyperbola should be shifted further from $(p_{1\parallel},p_{2\parallel})=(0,0)$ and the number of fringes increases, as indicated by Eq.~(\ref{Eq:fringes}).

 The remaining phase shifts, $\alpha_{lr}$ and $\alpha_{ld}$, will not vanish. However, by an adequate choice of parameters one may identify momentum regions in which they are smallest, which will give rise to interference maxima. The exponent $\alpha_{lu}$, which gives the interference between the left and the upper peaks,  behaves in a similar way as $\alpha_{ld}$, with the main difference that  Eq.~(\ref{phaseleftup}) shows an additional phase, with regard to Eq.~(\ref{phaseleftdown}) giving the left-down phase difference. This phase depends on $p_1^2+p_2^2$ and has a constant factor. Furthermore, the last term in Eq.~(\ref{phaseleftup}) causes a strong enhancement along the anti-diagonal.

\begin{figure}
\centering
\includegraphics[width=\linewidth]{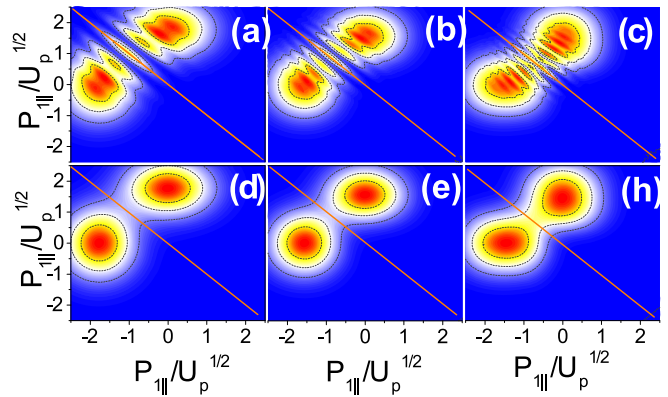}
\caption{Correlated electron-momentum distributions obtained by combining the transition amplitudes $M_l$ and $M_u$, isolating the effect of $\alpha_{l u}$, integrated over the transverse-momentum components. The driving-field parameters and ionization potentials are the same as in Fig.~\ref{fig:LeftandDownMaps}. We have also employed the same normalization and labeling as in Fig.~\ref{fig:LeftandDownMaps}, with the coherent and incoherent sums in the upper and lower panels, respectively. The anti-diagonals $p_{1\parallel}= -p_{2\parallel}$ are indicated with the orange lines in the figure.}
\label{fig:LeftandUpMaps}
\end{figure}

 The coherent and incoherent sums of $M_l$ and $M_u$ are presented in Fig.~\ref{fig:LeftandUpMaps}. The figure shows a very clear maximum along the anti-diagonal $p_{1\parallel}=-p_{2\parallel}$, and interference fringes with a richer substructure than the previous partial map.  These effects are caused by the additional phases mentioned above.  An estimate of the position of the fringes is not straightforward. However, we have verified that their spacing, for small momenta, is approximately one fourth of that given by Eq.~(\ref{Eq:fringes}). It also decreases with driving-field intensity.

\begin{figure}
\centering
\includegraphics[width=\linewidth]{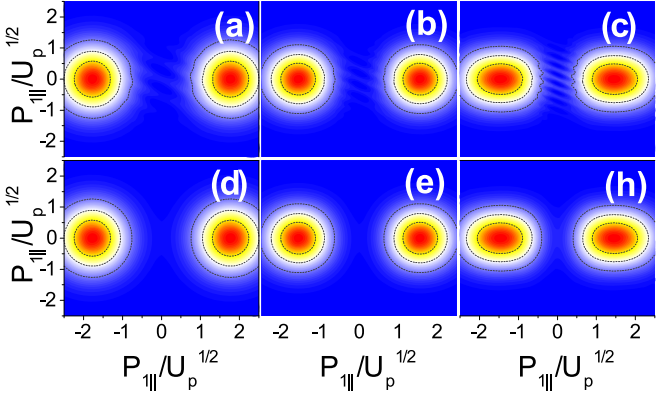}
\caption{Correlated electron-momentum distributions obtained by combining the transition amplitudes $M_l$ and $M_r$, isolating the effect of $\alpha_{l r}$ and integrating over the transverse momentum. The field parameters, ionization potentials and plotting style are the same as in Fig.~\ref{fig:LeftandDownMaps}.}
\label{fig:LeftandRightMaps}
\end{figure}

Finally, we display the partial sum between the left and right amplitudes $M_r$ and $M_l$ (Fig. \ref{fig:LeftandUpMaps}). In this case, the interference effects are minimal and only present close to the origin $(p_{1\parallel},p_{2\parallel})=(0,0)$. This is expected as according to the constraints the overlap between both amplitudes is vanishing in the other momentum regions. In the case of $\alpha_{lr}$ there is no obvious condition on the parallel momentum, independent of time or the perpendicular components, other than $p_{1 \parallel}$ and $p_{2 \parallel}$ being close or equal to zero. A time-dependent condition can be extracted,
\begin{align}
p_{2 \parallel}=-\frac{\sin(\omega t')}{\sin(\omega t)} p_{1 \parallel}. \label{eq:alpha1cond}
\end{align}
These trajectories will overlap for very low values of parallel momenta. In this case, for the dominant trajectories $t'$ is near a crossing and $t$ is near the next maximum \cite{Shaaran2010a}. These trajectories are located close to the axis. Near a crossing, $\sin(\omega t')\simeq \omega t'$ and near a maximum $\sin(\omega t)\simeq 1$. This strongly suggests that the slope in the overlap region will be constant as the rescattering time will not vary substantially. This is approximately the behavior observed in Fig.~\ref{fig:LeftandRightMaps}.  

\section{\label{sec4:level1}The Effect of the Prefactors}

Additionally to the interference effects studied above, the prefactors (\ref{eq:Vp1e,kg}) and (\ref{eq:Vp2e}) will introduce a momentum bias, which influences the shapes and, in principle, the quantum interference between events or channels. 
In the specific problem addressed in \cite{Hao2014}, the target chosen is Argon, whose first and second ionization potentials are $E_{1g}=0.58$ a.u. and $E_{2g}=1.02$ a.u., respectively. For the parameter range of interest, there exist six relevant excitation channels, which are provided in Table \ref{table:channels}
\begin{table}
	\begin{tabular}{c c c}
		\hline\hline
		Channel & Excited-State Configuration & $E_{2e}$ (a.u.)\\
		\hline
		1 & $3s3p^{6}$ ($3s \rightarrow 3p$ ) &0.52 \\
		2 & $3p^{5}3d$ ($3p \rightarrow 3d$) &0.41 \\
		3 & $3p^{5}4s$ ($3p \rightarrow 4s$) &0.4 \\
		4 & $3p^{5}4p$ ($3p \rightarrow 4p$) &0.31 \\
		5 & $3p^{5}4d$ ($3p \rightarrow 4d$) &0.18 \\
		6 & $3p^{5}5s$ ($3p \rightarrow 5s$) &0.19 \\
		\hline \hline
	\end{tabular}
	\caption{Relevant excitation channels for $Ar^+$, ordered according to principal and orbital quantum numbers for the second electron's excited state. From left to right, the columns give the number associated with the channel, the electronic configurations for the sub-levels involved in the excitation and the absolute value $E_{2e}$ of the excited-state energy, respectively. For clarity, the excitation pathway for 
	the second electron is given in brackets. }
	\label{table:channels}
\end{table}
and involve excitations to states of very different spatial geometry. Hence, they will give us a fairly good idea about the role of the prefactors. Throughout, we will restrict our studies to $m=0$, in order to facilitate a comparison with the results in \cite{Hao2014}.

In Eq.~(\ref{eq:excitationgeneral}) and (\ref{eq:ionizationgeneral}), we give the general expressions for the excitation and ionization prefactors, respectively, assuming that all bound-states involved are of the form $\psi_{nlm}(\mathbf{r})=R_{nl}(r)Y_l^{m}(\Omega)$, i.e., hydrogenic states. These prefactors have been first derived in \cite{Shaaran2010} in the context of a qualitative analysis, so that only their functional form has been emphasized. In the present work, we go beyond those qualitative expressions and include all normalization constants and phases, as they will be necessary for computing coherent superpositions.
\begin{widetext}
\begin{align}
\begin{split}
V_{p_1 e, k g}&=\sum_{L=|l_e-l_g|}^{l_e+l_g}\sum_{M=-L}^{L}(-i)^{L}A_1 Y^M_L(\theta_q,\phi_q)\frac{(\braket{l_g,l_e,0,0|L,0}\braket{lg,le,m_g,-m_e|L,M}}{\sqrt{(2L+1)}} I_r\\
&\\
I_r&=
\sum^{b_{n_gl_g}}_{k_g=0}\sum^{b_{n_el_e}}_{k_e=0}\frac{(-1)^{k_g+k_e}2^{a_1-1-2L}\xi^{L-a_1}\Xi^{n_g}_{l_gk_g}\Xi^{n_g}_{l_ek_e}\Gamma(a_1)}{k_g!\hspace{1mm}k_e!\hspace{1mm}(b_{n_gl_g})!\hspace{1mm}(b_{n_el_e})!\hspace{1mm}\Gamma(\frac{3}{2}+L)}d^{n_g}_{l_gk_g}d^{n_e}_{l_ek_e}
\left|\frac{q}{\xi}\right|^{L}{}_2F_1\left(\frac{1}{2}a_1,\frac{1}{2}(a_1+1);\frac{3}{2}+L;-\frac{q^2}{\xi^2}\right)\text{,}
\label{eq:excitationgeneral}
\end{split}
\end{align}
\end{widetext}
where
\begin{align*}
A_1&=(-1)^{m_e}C_{n_g l_g}C_{n_e l_e}&\frac{V_{12}(\bm{q})}{\sqrt{2\pi}}\sqrt{(2l_g+1)(2l_e+1)}
\end{align*}
\begin{align*}
C_{n l}&=\sqrt{\frac{(n-l-1)!}{2 n(n+l)!}}  &\Xi^n_{l k}&=\left(\sqrt{2 E_{n}}\right)^{\frac{3}{2}+l+k}\\
d^n_{lk}&=\frac{(n+l)!}{(2l+k+1)!} &\xi&=\sqrt{2 E_{n_g}}+\sqrt{2 E_{n_e}}\\
a_1&=3+k_g+k_e+l_g+l_e+L  &b_{n l}&=n-l-1\\
\end{align*}
Now here is the expression for $V_{p_2 e}$ again with all normalization constants  and phases.

\begin{align}
\begin{split}
V_{p_2, e}&=A_2\sum^{b_{n_el_e}}_{k=0} (-1)^{k} \frac{2^{k}(\sqrt{2 E_{n_e}})^{-\frac{1}{2}-l_e}p_2^{l_e}}
{(b_{n_el_e}-k)!\hspace{1mm}k!}d^{n_e}_{l_ek} \\
&\times\frac{\Gamma(a_2)}{\Gamma(\frac{3}{2}+l_e)}{}_{2}F_{1}\left(\frac{1}{2}a_2,\frac{1}{2}(a_2+1);\frac{3}{2}+l_e;-\frac{p_2^2}{2 E_{n_e}}\right)\text{.}
\intertext{Where,}
A_2&=2(-i)^{l_e} C_{n_e l_e} Y^{m_e}_{l_e}(\theta_{p_2},\phi_{p_2})\\
a_2 &= 2+k+2l_e\\
\label{eq:ionizationgeneral}
\end{split}
\end{align}

The above-stated prefactors have radial and angular nodes. For the first electron, Eq.~(\ref{eq:excitationgeneral}) depends on the intermediate momentum $\mathbf{k}(t'',t')$, which will vary with regard to $\mathbf{p}_1$. This will lead to these nodes being washed out to a great extent. We have verified that this happens even if the integration over $\mathbf{p}_{1\perp}$ is not performed. In general, transverse momentum integration will cause further blurring. Mostly, the prefactor $V_{\mathbf{p}_1e,\mathbf{k}g}$ will cause a shift in the peaks of the electron momentum distribution from $p_{1\parallel}=\pm 2 \sqrt{U_p}$ and alter their width.

The effects caused by the prefactor $V_{\mathbf{p}_2e}$ are much more dramatic.  This has been observed in our previous publications \cite{Shaaran2010,Shaaran2011c} for atoms and molecules, but has not been investigated systematically. Similarly to what is observed for hydrogenic wave functions, the number of radial nodes is given by $n_e-l_e-1$, and angular nodes by $l_e$. This is because, formally, the prefactor is the Fourier transform of a hydrogenic excited state $\psi_{n_el_em_e}(\mathbf{r}_2)$ modified by the interaction $V_{ion}(\mathbf{r}_2)=1/r_2$. Since $V_{ion}$ and $\exp(i \mathbf{p}_2) \cdot \mathbf{r}_2$ have no nodes, the number of nodes will be preserved but their energy positions will be different, if compared to the momentum-space wave function $\psi_{n_el_em_e}(\mathbf{p}_2)$. Their number and position with regard to the momentum $p_2=\sqrt{p^2_{2\parallel}+p^2_{2\perp}}$ are given in table \ref{tab:RadialNodes2}.

\begin{table*}
	\centering
	\begin{tabular}{|p{1.25cm}|p{0.75cm}|l|l|}
		\hline Channel and State & $b_{n_el_e}$ & Numerator & Roots \\ \hline
		1 $3p$ & 1 & $2 E_{2e}-p_2^2$ & $p_2=\sqrt{2 E_{2e}}$ \\
		2 $3d$ & 0 & const. & no roots\\
		3 $4s$ & 3 & $8 E_{2e}^3-28 E_{2e}^2 p_2^2+14 E_{2e} p_2^4-p_2^6$& $p_2=\sqrt{2 E_{2e}}$, $p_2 = \sqrt{(6\pm 4 \sqrt{2})E_{2e}}$ \\
		4 $4p$ & 2 & $20 E_{2e}^2-28 E_{2e} p_2^2+5 p_2^4$&$p_2= \sqrt{\frac{2}{5} \left(7\pm2 \sqrt{6}\right)E_{2e}} $\\
		5 $4d$ & 1 & $2 E_{2e}-p_2^2$ &$p_2=\sqrt{2 E_{2e}}$ \\
		6 $5s$ & 4 & $80 E_{2e}^4-480 E_{2e}^3 p_2^2+504 E_{2e}^2 p_2^4-120 E_{2e} p_2^6+5 p_2^8$ & $p_2= \sqrt{\left(2\pm\frac{4}{\sqrt{5}}\right)E_{2e}}$, $p_2= \sqrt{\left( 10\pm4 \sqrt{5}\right) E_{2e}} $\\
		\hline
	\end{tabular}
	\caption{Number of radial nodes $b_{n_el_e}=n_e-l_e-1$ and the numerator polynomials (and their associated roots) that give rise to these nodes. Note that $p_2^2=p^2_{2\parallel}+p^2_{2\perp}$, so that the expressions in the third column describe circles in the $p_{2\parallel}p_{2\perp}$ plane. }
	\label{tab:RadialNodes2}
\end{table*}

According to table \ref{tab:RadialNodes2}, the radial nodes will manifest themselves as circles in the $p_{2\parallel}p_{2\perp}$ plane. They are clearly seen if we fix the momentum of the first electron at $(p_{1\parallel},p_{1\perp})=(2\sqrt{U_p},0)$ and plot the probability distribution as a function of the momentum components $p_{2\parallel}$ and $p_{2\perp}$ of the second electron. This procedure is similar to the computation of partial momentum maps employed in our previous publications \cite{Shaaran2012,Faria2012}, and provide a wealth of detail which is lost if the transverse momentum integration is performed.

\begin{figure}
	\includegraphics[width=\linewidth]{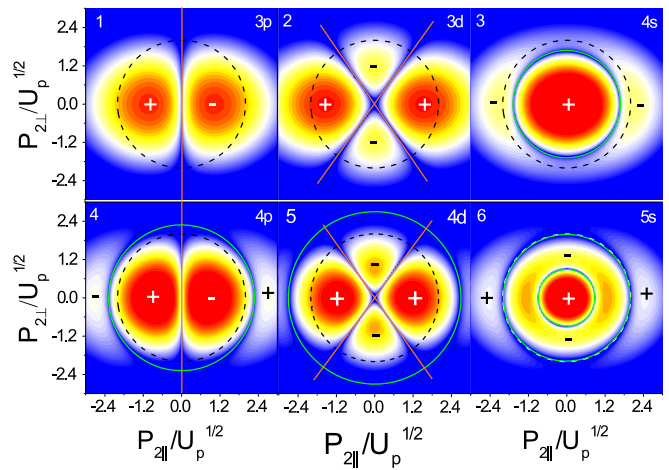}
	\caption{Cross-section of the total probability distribution with $p_1$ fixed at $(p_{1\parallel},p_{1\perp})=(2\sqrt{U_p},0)$, which gives an effective partial probability distribution over $p_2$. The ponderomotive energy is given by $U_p=0.1$ a.u.($I=4.56 \times 10^{13} \mathrm{W/cm}^2$). A logarithmic scale has been used to highlight the orbital-geometrical features. The radial and angular nodes resulting from the second ionisation prefactor are marked by green circles and red line respectively. The direct ATI cutoff $p^2_{2\parallel}+p^2_{2\perp}=4U_p$ is marked with a dashed circle. Beyond this point the probability distribution decays exponentially. Phases for each prefactor are indicated by + and - signs, with a change in sign indicating a flip. The signal in each panel has been normalized with regard to its maximum value. }
	\label{fig:partial}
\end{figure}

Fig.~\ref{fig:partial} displays these distributions for the six channels in Table \ref{table:channels}. The panel labels each correspond to the channel number, which is detailed in table \ref{table:channels}. The circle $p^2_{2\parallel}+p^2_{2\perp}=4U_p$ indicates the direct ATI cutoff, according to the condition (\ref{eq:constraint2nd}). Changes in the shapes of the distributions will be caused by nodes within this region. The radial nodes will then be particularly important for highly excited states, as in this case $E_{2e}$ is small. Physically, this is related to the fact that localization in momentum space corresponds to a position-space spread.
The smaller the binding energy, the more delocalized $\psi_{n_el_em_e}(\mathbf{r}_2)$ will be. 

The effect of the radial nodes can be seen by comparing channels 3 and 6, which involve $s$ states. For channel 3, there is only one radial node in the momentum region of interest, while for channel 6 the two existing nodes influence the electron momentum distributions. This will lead to an overall narrowing in momentum space. In the remaining channels, additionally to this effect, there are also angular nodes, which behave in very distinct ways. For $l_e=1$ (channels 1 and 4) they lead to a strong suppression in the electron-momentum distributions for $p_{2\parallel}=0$. Since these nodes occur for all $p_{2\perp}$, they will survive the transverse-momentum integration. This will cause the correlated two-electron distributions to move away from the axes.  For $d$ states, there are x-shaped nodes which intersect at $(p_{2\parallel},p_{2\parallel})=(0,0)$. We have verified that these nodes will also survive the integration over the transverse momentum components, but will lead to a secondary, much weaker maximum at the axes instead of a complete suppression. 

\begin{figure}
	\includegraphics[width=\linewidth]{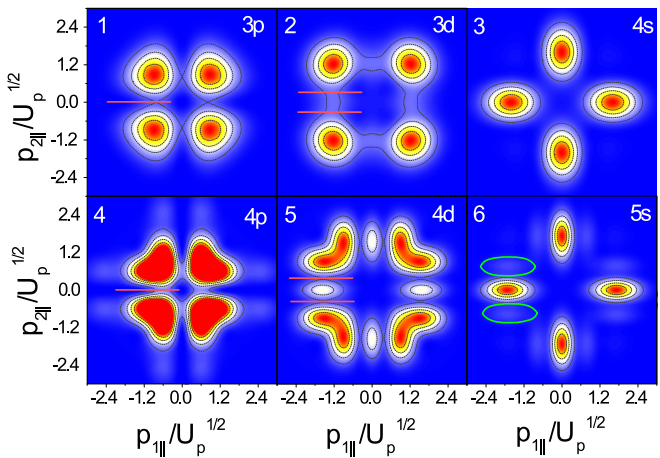}
	\caption{Full probability distribution with both prefactors included, an incoherent sum of events has been used. The panels are marked with the channel number and excitation state (top left and right corners, respectively). Red lines have been used to show the splitting caused by angular nodes and green circles mark the secondary peaks due to radial nodes. The ponderomotive energy is given by $U_p=0.1$ a.u.($I=4.56 \times 10^{13} \mathrm{W/cm}^2$). The yield in channel four has been over-exposed in order to show the secondary nodes.}
	\label{fig:fullprefactor1}
\end{figure}

\begin{figure}
	\includegraphics[width=\linewidth]{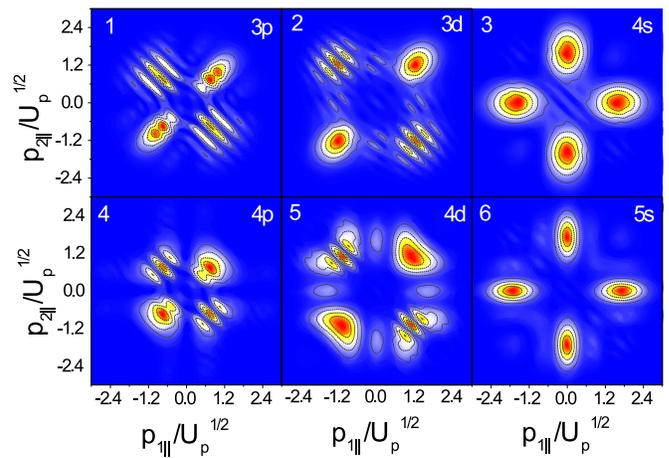}
	\caption{Full probability distribution with both prefactors included, for identical parameters as in figure \ref{fig:fullprefactor1}, except that a coherent sum of events has been used. The same panel labels are used as in Fig.~\ref{fig:fullprefactor1}. The signal in each panel has been normalized with regard to its maximum value.}
	\label{fig:fullprefactor2}
\end{figure}

In Fig.~\ref{fig:fullprefactor1}, we plot the incoherently symmetrized, correlated distributions, for the same channels as in Fig.~\ref{fig:partial}. The figure in fact shows an overwhelming influence of the prefactor $V_{\mathbf{p}_2e}$. Angular nodes in $p_{2\parallel}$ are clearly visible as cuts marked by orange lines and radial nodes can be seen by small secondary peaks marked by green circles. Only for very loosely bound states does the excitation prefactor lead to some substructure (see channel 4), although it is an order of magnitude below the main peak. This figure establishes which substructure comes from the prefactors themselves, so that they cannot be attributed to the interference between different events.

If a coherent sum is considered upon symmetrization, Fig.~ \ref{fig:fullprefactor2}, the same diagonal fringes can be seen as in Fig.~ \ref{fig:FullMaps}. Given interference only occurs along the diagonals, localization for $s$ states by the two prefactors, which narrows the distribution width and pushes the peak away from the origin, cause the diagonal and central region to be minimally occupied. Hence, little interference occurs for $s$ states. For $p$ and $d$ states, there is a lot of interference as the effect of the angular nodes is to split the distribution apart, widening it causing much of it to be along the diagonal. The actual type of interference is unchanged from figure \ref{fig:FullMaps}, we verified this by looking at the phase information from the prefactor. For the second ionization prefactor, looking a the phase plotted over $p_2$, the nodes represent a phase shifts of $\pi$, if this is applied to the partial momentum distribution there is little change in the resulting phase map and the effect of this after integration over $p_\perp$ will be lost entirely. Hence the prefactors effect the interference only by localization and all the effects derived, discussed earlier, are still valid.

\section{\label{sec5:level1}Interference of Channels}

We will now study the quantum interference between the different excitation channels in Table \ref{table:channels}. A uniform superposition of channels is used, which can be justified if one views each channel as a path the second electron can take from its ground state to the final Volkov state. Hence, the final transition amplitude should sum over the possible channels, leading to $|\sum_c M_c|^2$, where $M_c$ is the transition amplitude calculated for each channel.

\begin{figure}
	\includegraphics[width=\linewidth]{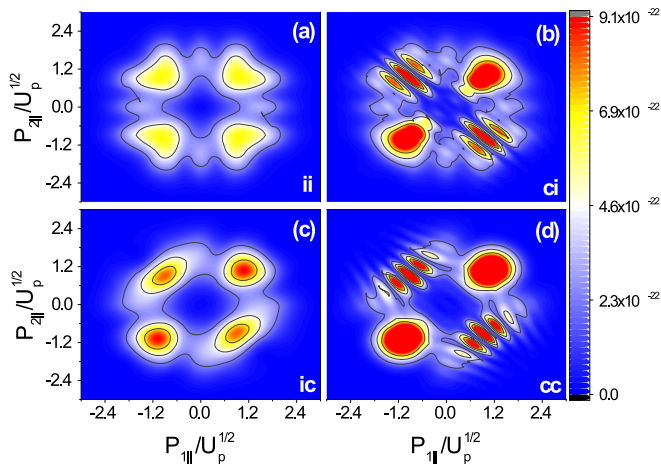}
	\caption{Full coherent and incoherent superpositions of all channels in Table \ref{table:channels}, for the same field parameters as in Fig.~\ref{fig:fullprefactor1}. Panel (a) is an incoherent sum of all the channels and their events, panel (b) is an incoherent sum of channels with a coherent sum of event, panel (c) is a coherent sum of channels with an incoherent sum of events and panel (d) is a coherent sum of channels and events. The symbols i and c in the bottom right corners denote incoherent and coherent sums for event and channel, respectively, with event preceding channel. All panels use the same arbitrary scale.}
	\label{fig:FullChanSum}
\end{figure}

In Fig.~\ref{fig:FullChanSum} we plot the full sum of channels 1 to 6, using different combinations of coherent and incoherent superpositions for events and channels. The figure shows that a fourfold momentum symmetry only occurs if the channels and events are summed incoherently [panel (a)]. Once quantum interference is introduced, only the reflection symmetry with regard to the diagonal or anti-diagonal remains, as shown in panels (b) to (d). In this case, the features along the diagonal and the anti-diagonal differ. However, only channel interference [panel (c)] exhibits a diagonal enhancement.  The anti-diagonal fringes only come from event interference [see panel (b)]. The diagonal enhancement and breaking of symmetry in panel (c) is consistent with what was found in \cite{Hao2014}.

\begin{table*}[t]
	\centering
	\begin{tabular}{|p{2.5cm}|p{2.5cm}|p{2.5cm}|}
		\hline
		$U_p=0.05$ & $U_p=0.1$ & $U_p=0.15$\\
		\hline
		\begin{tabular}{p{2mm}|c}
			1 & $3.49\times 10^{-28}$ \\
			4 & $2.02\times 10^{-28}$ \\
			5 & $1.54\times 10^{-28}$ \\
			2 & $1.01\times 10^{-28}$ \\
			3 & $7.78\times 10^{-29}$ \\
			6 & $6.17\times 10^{-30}$ \\
		\end{tabular}
		&
		\begin{tabular}{p{2mm}|c}
			5 & $2.95\times 10^{-22}$ \\
			4 & $1.31\times 10^{-22}$ \\
			1 & $9.91\times 10^{-23}$ \\
			2 & $7.62\times 10^{-23}$ \\
			3 & $4.94\times 10^{-23}$ \\
			6 & $1.21\times 10^{-23}$ \\
		\end{tabular}
		&
		\begin{tabular}{p{2mm}|c}
			5 & $4.67\times 10^{-19}$ \\
			4 & $1.02\times 10^{-19}$ \\
			2 & $5.43\times 10^{-20}$ \\
			1 & $3.26\times 10^{-20}$ \\
			3 & $2.52\times 10^{-20}$ \\
			6 & $1.94\times 10^{-20}$ \\
		\end{tabular}\\
		\hline
	\end{tabular}
	\caption{Mean values of the two electron parallel momentum probability distribution of each channel for different laser intensities, within the parameter range of interest. These distributions have been computed for a monochromatic field. }
	\label{tab:Dom}
\end{table*}
A legitimate question is whether one may identify dominant channels and/or features related to the channel type in the superpositions presented above. The shapes of the superpositions in \ref{fig:FullChanSum} suggest that excitations involving $p$ and $d$ states prevail.
Table \ref{tab:Dom} shows the mean values of the correlated electron momentum distributions for each channel, which are comparable. Since one channel does not dominate significantly over the rest, interference is expected to be important.  This is contrary to the results in \cite{Hao2014}, where channels 1-3 were found to dominate.

\begin{figure}
	\includegraphics[width=\linewidth]{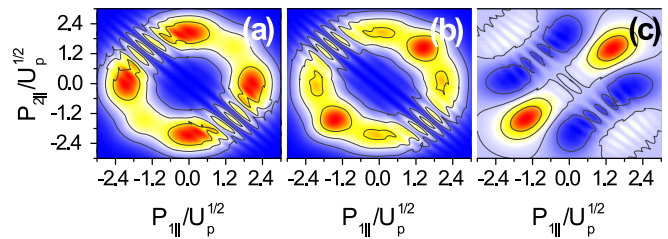}
	\caption{Channel sum 1 and 4 without prefactors. Panel a) shows an incoherent sum of channel and a coherent sum of events, panel b) shows a coherent sum of channels and events, panel c) is the difference between the two. The driving-field parameters are the same in Fig.~\ref{fig:fullprefactor1}. The signal in each panel has been normalized to its maximum value.}
	\label{fig:ChanSum0}
\end{figure}

More insight is obtained by considering superpositions of two channels, which may be incoherent or coherent. The former and the latter case are given by
\begin{equation}
\Omega_{In}(p_{1 \parallel},p_{2 \parallel})=||M_1||^2 +||M_2||^2\label{Eq:incoh}
\end{equation}
and
\begin{equation}
\Omega_{Coh}(p_{1 \parallel},p_{2 \parallel})=||M_1 + \text{e}^{i \phi} M_2||^2, \label{Eq:coh}
\end{equation}
respectively. In the coherent sum (\ref{Eq:coh}), we have included a phase $\phi$ that can be used to manipulate interference effects such as diagonal or anti-diagonal enhancement.

Without the effect of prefactors there is little qualitative difference between the possible channel sums, given that the actions only differ by the term $E_{2 g} t$ [see panels (a) and (b) in Fig.~\ref{fig:ChanSum0}]. Nonetheless, in the difference between the coherent and incoherent sums we can see hyperbolic fringes [Fig.~\ref{fig:ChanSum0}(c)]. We have verified empirically that the position of the fringes is determined by the value of the phase $\phi$ and the thickness of the fringes is inversely related to $E_{2 e_1}-E_{2 e_2}$. Due to their location in the $p_{2\parallel}p_{1\parallel}$ plane, the most significant interfering terms will be equivalent events between channels, e.g., $M_{1l}$ and $M_{2l}$, but the terms related by particle exchange such as $M_{1l}$ and $M_{2d}$ will also be important. The prefactors add a momentum-dependent phase difference between the two channels. In the case of the second ionization prefactor, which mainly determines the interference effects, the phase is constant but inverts when a nodal line is crossed. The other prefactors depend on $\mathbf{k}$, which has a complex phase relation determined by the saddle point equations.

\begin{figure}
	\includegraphics[width=\linewidth]{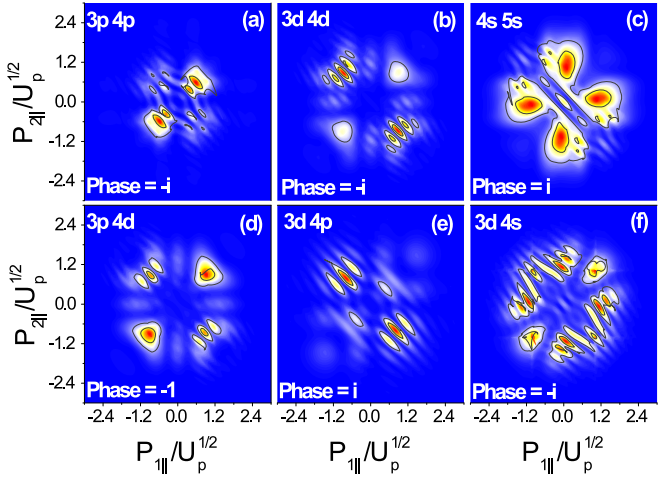}
	\caption{Two-channel sums with prefactors, for the same driving-field parameters in Fig.~\ref{fig:fullprefactor1}. The numbers at the top right in each panel labels the excited states used in the superposition, with the phase difference given in the bottom left. The signal in each panel has been normalized to its maximum value. }
	\label{fig:ChanSum2}
\end{figure}

Figure \ref{fig:ChanSum2} shows a selection of particular interference phenomena occurring in two-channel sums. In panel (a) the recollision and second ionization prefactor both cause phase inversion, which overall cancels. The diagonal enhancement comes from the fringes related to channel interference shifted by a phase of $-i$, whereas $i$ would cause an anti-diagonal enhancement.  For panel (b) the total effect of the prefactors is to cause an inversion in the channel-interference fringes. This leads to a suppressed signal along the diagonal. However, the effect is not as strong as the fringes are distorted by $V_{\mathbf{p}_2e}$. In both panels (a) and (b), the thickness of the channel-interference fringes are comparable to those associated with the ``left-down" interference. There are no significant diagonal effects in panel (c) as both distributions are concentrated near the axes $p_{n\parallel}=0 (n=1,2)$. This happens because, for the two interfering channels, the second electron is excited to an $s$ state. However, there are some interference effects breaking the fourfold momentum symmetry.

The remaining panels show some implications of channel interference involving energetically very close and distant excited states. For the interference of channels 1 and 5, shown in panel (d), there is a large difference in the excited-state energies. This causes small inter-channel fringes, so that suppression along the diagonal or anti-diagonal is not possible. In contrast, for panel (e), the excited bound states are energetically very close. This implies that fringes stemming from channel interference are too thick to cause a diagonal or anti-diagonal suppression. However, the prefactor does this instead (see below). For panel (f) the channel-related fringes are even thicker, so that the substructure is determined by the event interference and the prefactors. 

\begin{figure}
	\includegraphics[width=\linewidth]{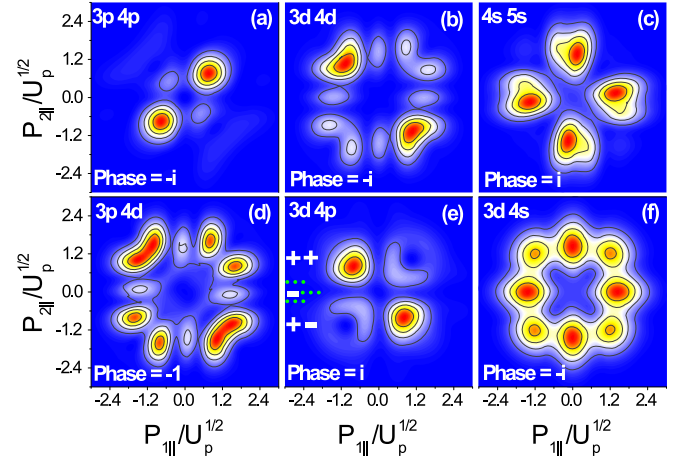}
	\caption{Same two-channel sums as Figure \ref{fig:ChanSum2}, except that the events have been summed incoherently. For clarity, the prefactor phase information has been marked on panel (e) in white. This information can be related to panels 2 and 4 in Fig.~\ref{fig:partial}. The green dotted lines mark nodal lines. The same logic can be applied to panel (f), with the difference that the $4s$ state does not have angular nodes.}
	\label{fig:ChanSumInCoh}
\end{figure}

Figure \ref{fig:ChanSumInCoh} shows the same coherent sums of channels as in Fig.~\ref{fig:ChanSum2}, but, instead, incoherent sums of events. All the diagonal and anti-diagonal effects remain and are in general stronger without the phase information and fringes from different events. The influence of the combined prefactors can also be seen more clearly. For instance, in panels (a) to (c) the features related to  $p$, $d$ and $s$ states are very evident, with a further bias introduced by the inter-channel interference. These features are (a) probability densities concentrated at the diagonal and anti-diagonal; (b) similar probability distributions as in (a), but with secondary maxima at the axes; (c) distributions concentrated mostly at the axes, respectively. This happens because the angular momentum quantum number $l_e$ of the excited states are the same for the two interfering channels.
 
The situation becomes more complex in the lower panels, in which channels with different angular momenta $l_e$ are mixed. In this case, we have identified two very striking scenarios, which occur for energetically close levels [panels (e) and (f)]. Because the inter-channel fringes are very thick in these cases, the shape of the electron-momentum distributions will be mainly determined by the prefactors and their phases. For instance, the clear diagonal suppression for panel (e) can be explained by the phase of the second ionization prefactors, which flips at every nodal line. The interfering channels involve excitation to $3d$ and $4p$. For a $p$ state, there is one node at $p_{2\parallel}=0$, while for a $d$ state there are two. This means that, from the top to  the  bottom of the panel, the phase of the $p$ state will flip once, while that of the $d$ state will flip twice. Hence, in the second quadrant the channels interfere constructively, while in the third quadrant destructive interference occurs. The same line of argument can be employed for the first and fourth quadrant, but in this case the interference pattern will be reversed.  This shows a case where we have entirely prefactor dependent anti-diagonal enhancement. In panel (f), the phase of the $4s$ prefactor will not flip in the momentum region of interest, while that of the $3d$ prefactor will flip twice. Hence, this will preserve the fourfold symmetry. Furthermore, interference between the channels will not be significant, as d states populate mainly the two diagonals and s states lead to distribution localized along the axes. This leads to a momentum distribution with peaks at the axes and the two diagonals. This distribution is, for practical purposes, fourfold symmetric, unless event interference is considered [see Fig.~\ref{fig:ChanSum2}(f)].

To summarize the inter-channel fringes, prefactors and a relative phase can cause a range of interference effects. The prefactors can cause an inversion, which leads to diagonal/anti-diagonal enhancement being swapped. They can also apply a phase shift, such that a different phase between the two channels is needed for diagonal/ anti-diagonal enhancement. Inter-channel interference effects are not washed out by more complex superpositions, as can be seen by directly comparing Figs.~\ref{fig:FullChanSum}(a) and (c).
\section{\label{sec6:level1}Discussion and Conclusions}

In this work, we have performed an in-depth, semi-analytical study of quantum interference in recollision-excitation with subsequent ionization (RESI) using the strong-field approximation (SFA). Our analysis includes interference of symmetry-related features such as electron indistinguishability, and of different excitation channels. Overall, we have found that the electron momentum distributions are greatly influenced by both types of interference.  The main effect of quantum interference is to break their fourfold symmetry in $p_{1\parallel}p_{2\parallel}$ plane, while the symmetry with regard to the diagonals is retained. This fourfold symmetry has been encountered in previous RESI studies employing the strong-field approximation \cite{Shaaran2010,Shaaran2010a,Shaaran2011} or related methods \cite{Chen2010}.

 We have shown, by considering a coherent sum of symmetry-related events, that interference effects previously thought to be washed out from integration over perpendicular momenta are present in the correlated electron-momentum distributions. These effects affect the RESI distributions already for a single channel of excitation, via enhancement and suppression near the diagonals $p_{1\parallel}=\pm p_{2\parallel}$. We provide fully analytical expressions and estimates for diagonal, anti-diagonal and hyperbolic interference patterns. Similar fringes can be seen, but have not been explained, in \cite{Wang2012}, in which RESI has been modeled using a strong-field quantum-electrodynamical method (see Fig.~7 therein).
 
 We have also found that inter-channel interference will play an important role in RESI, in agreement with the results of \cite{Hao2014}. We go, however, beyond such studies and show that the shape of the electron-momentum distributions will be determined by a complex interplay of inter-channel and event interference, and the geometry of the excited bound states. This will mainly occur near the diagonal and anti-diagonal in the parallel-momentum plane. This means that it will mainly affect channels involving excitation to $p$ or $d$ states, while the influence on those with $s$-state excitation  will be much less critical. In particular, for the parameter range of interest, the contributions from all channels used in this work are comparable.
 
 We also analyze this interference in more detail using two-channel coherent superpositions. In this case, diagonal or anti-diagonal enhancement may occur due to inter-channel fringes and/or geometry-dependent prefactors. The fringes have hyperbolic shape, and their width is inversely proportional to the energy difference between the two channels involved. The prefactors will determine the region in momentum space to be occupied. In particular the nodes of the ionization prefactor $V_{\mathbf{p}_{2}e}$ of the second electron will cause phase shifts, which will influence inter-channel interference in specific momentum ranges. This enhancement can be manipulated using a relative phase. In this context, one should notice that by appropriate choice of channels and relative phase one may obtain anti-correlated distributions without resorting to bound-state depletion. This latter feature has been employed in \cite{Hao2014} in order to suppress the signal in the first and third quadrants of the parallel-momentum plane.
 
 Interestingly, depending on how interference occurs, the RESI distributions may exhibit diagonal enhancement (correlation), anti-diagonal enhancement (anti-correlation), or be spread in the four quadrants of the $p_{1\parallel}p_{2\parallel}$ plane. In contrast, for electron-impact ionization, the probability density is located only in the first and third quadrants and interference effects get washed out by transverse momentum integration. This sheds some light on experimental findings where different atoms give either diagonal or anti-diagonal enhancement \cite{Liu2010}.  Diagonal enhancement is normally attributed to electron-impact ionization. However, for low, below-threshold intensities, this could also be related to RESI. Indeed, all possible results found experimentally \cite{Eremina2003, Liu2008, Liu2010} are achievable if we can find the correct superposition of channels. Furthermore there exist theoretical studies for which anti-correlation has been obtained without excitation \cite{Bondar2009}. This suggests that the ability to manipulate diagonal and anti-diagonal enhancement with a phase opens up the possibility of control over the RESI process, which could lead to various applications. 
 
Finally, we would like to comment on quantum-classical correspondence in RESI. There has been considerable debate in the literature whether NSDI in general and RESI in particular is a classical or quantum mechanical phenomenon. On the one hand, our results show that quantum interference has a striking influence on the shapes and localization of the electron-momentum distributions. Hence, classic-trajectory methods must be viewed with care. On the other hand, highly excited states may give rise to a quasi-continuum, which would allow the existence of a quasi-classical wave packet. This would justify the success of classical models. For molecules, a larger density of states and enhanced ionization may increase their predictive power \cite{Ye2008a,Li2014,Emmanouilidou2009}. 

Furthermore, the SFA considers discrete states and neglects broadening and distortion caused by the field. It could well be that these effects lead to a strong overlap and thus the creation of a quasi-continuum, washing out phase information. However, recent studies of the RESI dynamics in phase space have revealed a highly confined region that can be associated with trapping in an excited state \cite{Mauger2012}. This would justify using discrete bound states and neglecting depletion, and would render interference important. Another feature which has not been included is the influence of the Coulomb potential, which changes the topology of the electron trajectories \cite{Yan2010,Lai2015}. For a detailed discussion on the advantages and drawbacks of classical and quantum-mechanical approaches in the modeling of RESI see our review article \cite{Faria2011}.  The present work contributes to this discussion by shedding additional light on the role of interference in this context.

\section*{Acknowledgements}
We would like to thank J. Chen and X. Liu for useful discussions, and the Chinese Academy of Sciences, Wuhan for its kind hospitality. We would also like to thank Joey Dumont helping us with his complex Bessel function library \url{https://github.com/valandil/complex_bessel}. The authors acknowledge the use of the UCL Legion High- Performance Computing Facility (Legion@UCL), and associated support services, in the completion of this work. This work was funded by the UK EPSRC (Engineering and Physical Sciences Research Council)(EP/J019240/1 and Doctoral Training Account).


\begin{thebibliography}{10}

\bibitem{Lein2007}
M. Lein, J. Phys. B: At. Mol. Opt. Phys. {\bf 40},  R135  (2007).

\bibitem{Augstein2012}
B. Augstein and C. {Figueira de Morisson Faria}, Mod. Phys. Lett. B {\bf 26},
  1130002  (2012).

\bibitem{Rudenko2004a}
A. Rudenko {\it et~al.}, J. Phys. B: At. Mol. Opt. Phys. {\bf 37},  L407
  (2004).

\bibitem{Chen2006}
Z. Chen {\it et~al.}, Phys. Rev. A {\bf 74},  053405  (2006).

\bibitem{Arbo2006a}
D.~G. Arb\'{o} {\it et~al.}, Phys. Rev. Lett. {\bf 96},  143003  (2006).

\bibitem{Arbo2008}
D.~G. Arb\'{o} {\it et~al.}, Phys. Rev. A {\bf 77},  013401  (2008).

\bibitem{Yan2012}
T.-M. Yan and D. Bauer, Phys. Rev. A {\bf 86},  053403  (2012).

\bibitem{Corkum1993}
P. Corkum, Phys. Rev. Lett. {\bf 71},  1994  (1993).

\bibitem{Faria2011}
C. {Figueira de Morisson Faria} and X. Liu, J. Mod. Opt. {\bf 58},  1076
  (2011).

\bibitem{Becker2012}
W. Becker, X. Liu, P.~J. Ho, and J.~H. Eberly, Rev. Mod. Phys. {\bf 84},  1011
  (2012).

\bibitem{Ye2008}
D.~F. Ye, X. Liu, and J. Liu, Phys. Rev. Lett. {\bf 101},  233003  (2008).

\bibitem{Emmanouilidou2008}
A. Emmanouilidou, Phys. Rev. A {\bf 78},  23411  (2008).

\bibitem{Panfili2001}
R. Panfili, J.~H. Eberly, and S.~L. Haan, Opt. Express {\bf 8},  431  (2001).

\bibitem{FigueiradeMorissonFaria2004a}
C. {Figueira de Morisson Faria}, H. Schomerus, X. Liu, and W. Becker, Phys.
  Rev. A {\bf 69},  043405  (2004).

\bibitem{Faria2004}
C. {Figueira de Morisson Faria}, X. Liu, A. Sanpera, and M. Lewenstein, Phys.
  Rev. A {\bf 70},  043406  (2004).

\bibitem{Jia2013}
X. Jia {\it et~al.}, Phys. Rev. A {\bf 88},  033402  (2013).

\bibitem{Weber2000}
T. Weber {\it et~al.}, Nature {\bf 405},  658  (2000).

\bibitem{Eremina2003}
E. Eremina {\it et~al.}, J. Phys. B: At. Mol. Opt. Phys. {\bf 36},  3269
  (2003).

\bibitem{Zeidler2005}
D. Zeidler {\it et~al.}, Phys. Rev. Lett. {\bf 95},  203003  (2005).

\bibitem{Liu2008}
Y. Liu {\it et~al.}, Phys. Rev. Lett. {\bf 101},  053001  (2008).

\bibitem{Liu2010}
Y. Liu {\it et~al.}, Phys. Rev. Lett. {\bf 104},  173002  (2010).

\bibitem{Bergues2012}
B. Bergues {\it et~al.}, Nat. Commun. {\bf 3},  813  (2012).

\bibitem{Kubel2014}
M. K\"{u}bel {\it et~al.}, New J. Phys. {\bf 16},  033008  (2014).

\bibitem{Sun2014}
X. Sun {\it et~al.}, Phys. Rev. Lett. {\bf 113},  103001  (2014).

\bibitem{Shaaran2010}
T. Shaaran, M.~T. Nygren, and C. {Figueira de Morisson Faria}, Phys. Rev. A
  {\bf 81},  063413  (2010).

\bibitem{Shaaran2010a}
T. Shaaran and C. {Figueira de Morisson Faria}, J. Mod. Opt. {\bf 57},  984
  (2010).

\bibitem{Chen2010}
Z. Chen, Y. Liang, and C.~D. Lin, Phys. Rev. A {\bf 82},  063417  (2010).

\bibitem{Emmanouilidou2009}
A. Emmanouilidou and A. Staudte, Phys. Rev. A {\bf 80},  053415  (2009).

\bibitem{Ye2010}
D.~F. Ye and J. Liu, Phys. Rev. A {\bf 81},  043402  (2010).

\bibitem{Zhang2014}
L. Zhang {\it et~al.}, Phys. Rev. A {\bf 90},  061401  (2014).

\bibitem{Shaaran2011}
T. Shaaran, B.~B. Augstein, and C. {Figueira de Morisson Faria}, Phys. Rev. A
  {\bf 84},  013429  (2011).

\bibitem{Hao2014}
X. Hao {\it et~al.}, Phys. Rev. Lett. {\bf 112},  073002  (2014).

\bibitem{Shaaran2012}
T. Shaaran, C. {Figueira de Morisson Faria}, and H. Schomerus, Phys. Rev. A
  {\bf 85},  023423  (2012).

\bibitem{Faria2002}
C. {Figueira de Morisson Faria}, H. Schomerus, and W. Becker, Phys. Rev. A {\bf
  66},  043413  (2002).

\bibitem{Faria2012}
C. {Figueira de Morisson Faria}, T. Shaaran, and M.~T. Nygren, Phys. Rev. A
  {\bf 86},  053405  (2012).

\bibitem{Shaaran2011c}
T. Shaaran, B.~B. Augstein, and C. {Figueira de Morisson Faria}, Phys. Rev. A
  {\bf 84},  013429  (2011).

\bibitem{Wang2012}
B. Wang {\it et~al.}, Phys. Rev. A {\bf 85},  023402  (2012).

\bibitem{Bondar2009}
D.~I. Bondar, W.-K. Liu, and M.~Y. Ivanov, Phys. Rev. A {\bf 79},  023417
  (2009).

\bibitem{Ye2008a}
D.~F. Ye, J. Chen, and J. Liu, Phys. Rev. A {\bf 77},  013403  (2008).

\bibitem{Li2014}
Y. Li, S.~P. Yang, J. Chen, and J. Fan, J. Phys. B: At. Mol. Opt. Phys. {\bf
  47},  045601  (2014).

\bibitem{Mauger2012}
F. Mauger, A. Kamor, C. Chandre, and T. Uzer, Phys. Rev. Lett. {\bf 108},
  063001  (2012).

\bibitem{Yan2010}
T.-M. Yan, S.~V. Popruzhenko, M.~J.~J. Vrakking, and D. Bauer, Phys. Rev. Lett.
  {\bf 105},  253002  (2010).

\bibitem{Lai2015}
X.~Y. Lai, C. Poli, H. Schomerus, and C. {Figueira de Morisson Faria},
  arXiv:1506.03646  16  (2015).

\end{thebibliography}
\end{document}